\documentclass[letter,traditabstract]{aa} % for the letters 
\usepackage{graphicx}
\usepackage{natbib}
\usepackage{lscape}
\usepackage{longtable}
\usepackage{txfonts}
\usepackage{caption}
\usepackage{subcaption}

\usepackage{float}

\def\mstar  {$M_{\star}$}
\def\macc   {$\dot{M}_{\rm acc}$}
\def\mdisk {$M_{\rm disk}$}
\def\mdust {$M_{\rm disk,dust}$}
\def\mgas {$M_{\rm disk,gas}$}
\def\lacc   {$L_{\rm acc}$}

\def\msun {$M_{\odot}$}

\defcitealias{Ansdell16}{AW16}

\begin{document}

   \title{Evidence for a correlation between mass accretion rates \\onto young stars and the mass of their protoplanetary disks}

    \author{C.~F. Manara \inst{1}\fnmsep\thanks{ESA Research Fellow}, G. Rosotti\inst{2}, L. Testi\inst{3,4,5}, A. Natta\inst{4,6}, J.~M. Alcal\'a\inst{7}, J.~P. Williams\inst{8}, M. Ansdell\inst{8}, A. Miotello\inst{9}, N. van der Marel\inst{8,9}, M. Tazzari\inst{3,5}, J. Carpenter\inst{10}, G. Guidi\inst{4}, G.~S. Mathews\inst{11}, I. Oliveira\inst{12}, T. Prusti\inst{1}, E.~F. van Dishoeck\inst{9,13}
          }

\titlerunning{Macc vs Mdisk}

\authorrunning{Manara et al.}

   \institute{Scientific Support Office, Directorate of Science, European Space Research and Technology Centre (ESA/ESTEC), Keplerlaan 1, 2201 AZ Noordwijk, The Netherlands \\
              \email{cmanara@cosmos.esa.int}
\and
Institute of Astronomy, University of Cambridge, Madingley Road, Cambridge CB30HA, UK
\and
European Southern Observatory, Karl-Schwarzschild-Str. 2, D-85748 Garching bei M\"unchen, Germany
\and
INAF/Osservatorio Astrofisico di Arcetri, Largo E. Fermi 5, I-50125 Firenze, Italy
\and
Excellence Cluster Universe, Boltzmannstr. 2, 85748 Garching bei M\"unchen, Germany
\and
School of Cosmic Physics, Dublin Institute for Advanced Studies, 31 Fitzwilliams Place, 2 Dublin, Ireland
\and
INAF/Osservatorio Astronomico di Capodimonte, Salita Moiariello, 16 80131 Napoli, Italy
\and
Institute for Astronomy, University of Hawai'i at M\"anoa, Honolulu, HI, USA
\and
Leiden Observatory, Leiden University, PO Box 9513, 2300 RA Leiden, The Netherlands
\and
California Institute of Technology, 1200 East California Blvd, Pasadena, CA 91125, USA
\and
University of Hawaii, Department of Physics and Astronomy, 2505 Correa Rd. Honolulu, Hawaii 96822 
\and
Observat\'orio Nacional/MCTI, Rio de Janeiro, 20921-400, Brazil
      \and
Max-Plank-Institut f\"ur Extraterrestrische Physik, Giessenbachstra{\ss}e 1, D-85748 Garching, Germany
             }

  \date{Received March 18, 2016; accepted May, 18th 2016}

% \abstract{}{}{}{}{} 
% 5 {} token are mandatory
 
  \abstract
{
A relation between the mass accretion rate onto the central young star and the mass of the surrounding protoplanetary disk has long been theoretically predicted and observationally sought. 
For the first time, we have accurately and homogeneously determined the photospheric parameters, mass accretion rate, and disk mass for an essentially complete sample of young stars with disks in the Lupus clouds. Our work combines the results of surveys conducted with VLT/X-Shooter and ALMA.
With this dataset we are able to test a basic prediction of viscous accretion theory, the existence of a linear relation between the mass accretion rate onto the central star and the total disk mass. 
We find a correlation between the mass accretion rate and the disk dust mass, with a ratio that is roughly consistent with the expected viscous timescale when assuming an interstellar medium (ISM) gas-to-dust ratio. 
This confirms that mass accretion rates are related to 
the properties of the outer disk. 
We find no correlation between mass accretion rates and the disk mass measured by CO isotopologues emission lines, possibly owing to the small number of measured disk gas masses. This suggests that the mm-sized dust mass better traces the total disk mass and that masses derived from CO may be underestimated, at least in some cases.

}
%   {TBW}
%  % aims heading (mandatory)
%   {TBW.}
%  % methods heading (mandatory)
%   {TBW}
%  % results heading (mandatory)
%   {TBW}
%  % conclusions heading (optional), leave it empty if necessary 
%   {}


   \maketitle
%
%________________________________________________________________

\section{Introduction}

The evolution of a protoplanetary disk significantly influences the planetary system that is formed. The final mass distribution of planets resembles the evolution of the surface density of gas in the disk \citep[e.g.,][]{Thommes08} and more massive disks lead to systems with more massive planets \citep{Mordasini12}. The evolution of the disk structure is mainly driven by processes happening in the disk, such as dust evolution \citep{TestiPPVI}, and by interaction between the disk and the central star through viscous accretion and winds \citep{AlexanderPPVI}. 

In the context of viscously evolving protoplanetary disks, 
the mass accretion rate onto the central star (\macc) and the mass of the disk (\mdisk) should be directly correlated \citep[e.g., Eq.~7 of][]{Hartmann98}. The ratio between these quantities is related to the viscous timescale ($t_\nu$) at the outer radius of the disk $(R_{\rm out})$ and the assumptions about the disk viscous properties. Overall, it is expected that \macc $\sim$ \mdisk/$t_\nu(R_{\rm out})$ with a coefficient of order unity \citep[e.g.,][]{Jones12}. In disks that evolved viscously, the ratio \mdisk/\macc \ must be comparable to the age of the system independent of the initial conditions and value of the viscosity parameter $\alpha$. The tight correlation between \macc \ and stellar masses \citep[\macc$\propto$\mstar$^{1.8}$, e.g.,][]{Muzerolle03,Natta06,Alcala14,Manara16} was also explained by \citet{Dullemond06} as a consequence of the initial rotation rate of the cores where disks formed. They predict a strong dependence of \mdisk \ on $M_\star^2$ and a tight correlation of \mdisk \ with \macc \ as a consequence of viscous evolution. This theoretical relation between \mdisk \ and \macc \ has been empirically investigated, but previous studies were unable to find any significant correlation \citep[e.g.,][]{Andrews10,Ricci10}.

In this Letter we present a study of an almost complete and homogeneous dataset of young stars in the $\sim$1-3 Myr old Lupus star-forming region \citep[][$d$=150-200 pc]{Comeron08}. We collected 
\macc \ measured from ultraviolet (UV) excess with the VLT/X-Shooter spectrograph, and \mdisk \ measured both from sub-mm continuum and CO line emission with Atacama Large Millimeter/submillimeter Array (ALMA). We look for correlations between \macc \ and \mdisk,
as predicted by viscous theory. 

%__________________________________________________________________

\section{Data sample}\label{sect::sample}

The sample analyzed here includes Class~II and transition disk (TD) young stellar objects (YSOs) with 0.1$<$\mstar/\msun$<$2.2, thus including only the TTauri stars of the ALMA sample.
Both the ALMA and X-Shooter surveys are complete at the $\sim$95\% level. In total, there are 66 objects with ALMA  and X-Shooter data available. The list of targets included in the analysis is reported in Table~\ref{tab::sample1}.

The ALMA data are presented by \citet[][hereafter AW16]{Ansdell16}. The setting includes continuum emission at 335.8 GHz (890 $\mu$m) at a resolution of $\sim$0.34\arcsec$\times$0.28\arcsec ($\sim$25$\times$20 AU radius at 150 pc) and the $^{13}$CO and C$^{18}$O 3-2 transitions. From the continuum emission, detected for 54 of the targets included here, \citetalias{Ansdell16} derive disk dust mass (\mdust) using typical assumptions of a single dust grain opacity $\kappa$(890 $\mu$m) = 3.37 cm$^2$/g
and a single dust temperature $T_{\rm dust}$=20 K. 
From the CO emission lines, \citetalias{Ansdell16} derive the disk gas mass (\mgas) for 29 disks in our sample, but for 22 of these the lower bound of the acceptable values of \mgas \ is unconstrained because of the lack of C$^{18}$O detection. Upper limits are calculated for the other 37 targets.

We obtain \macc from the X-Shooter spectra\ \citep{Alcala14, Alcala16}. Briefly, the stellar and accretion parameters are derived by finding the best fit among a grid of models including photospheric templates, a slab model for the accretion spectrum, and reddening. We use the UV excess as a main tracer of accretion and the broad wavelength range covered by X-Shooter ($\lambda\lambda \sim 330-2500$) to constrain both the spectral type of the target and the extinction. Among the objects discussed here, 57 have \macc \ derived from X-Shooter measurements, while 5 have an accretion rate compatible with chromospheric noise (non accretors), and 4 targets are observed edge-on, thus their \macc \ are underestimated. The evolutionary models by \citet{Siess00} are used to determine \mstar and, thus, \macc.

Finally, the sample includes several resolved binaries. All of these binaries have separations $\gtrsim2\arcsec$ and are indicated in Table~\ref{tab::sample1}. 

%__________________________________________________________________

\section{Results}

%__________________________________________________________________

\subsection{Disk dust mass}\label{sect::obs_macc_mdust}

%%%%%%%%%%%%%%%%%%%%%%%%%%%%%%%%%%%%%
\begin{figure}[!t]
\centering
\includegraphics[width=0.5\textwidth]{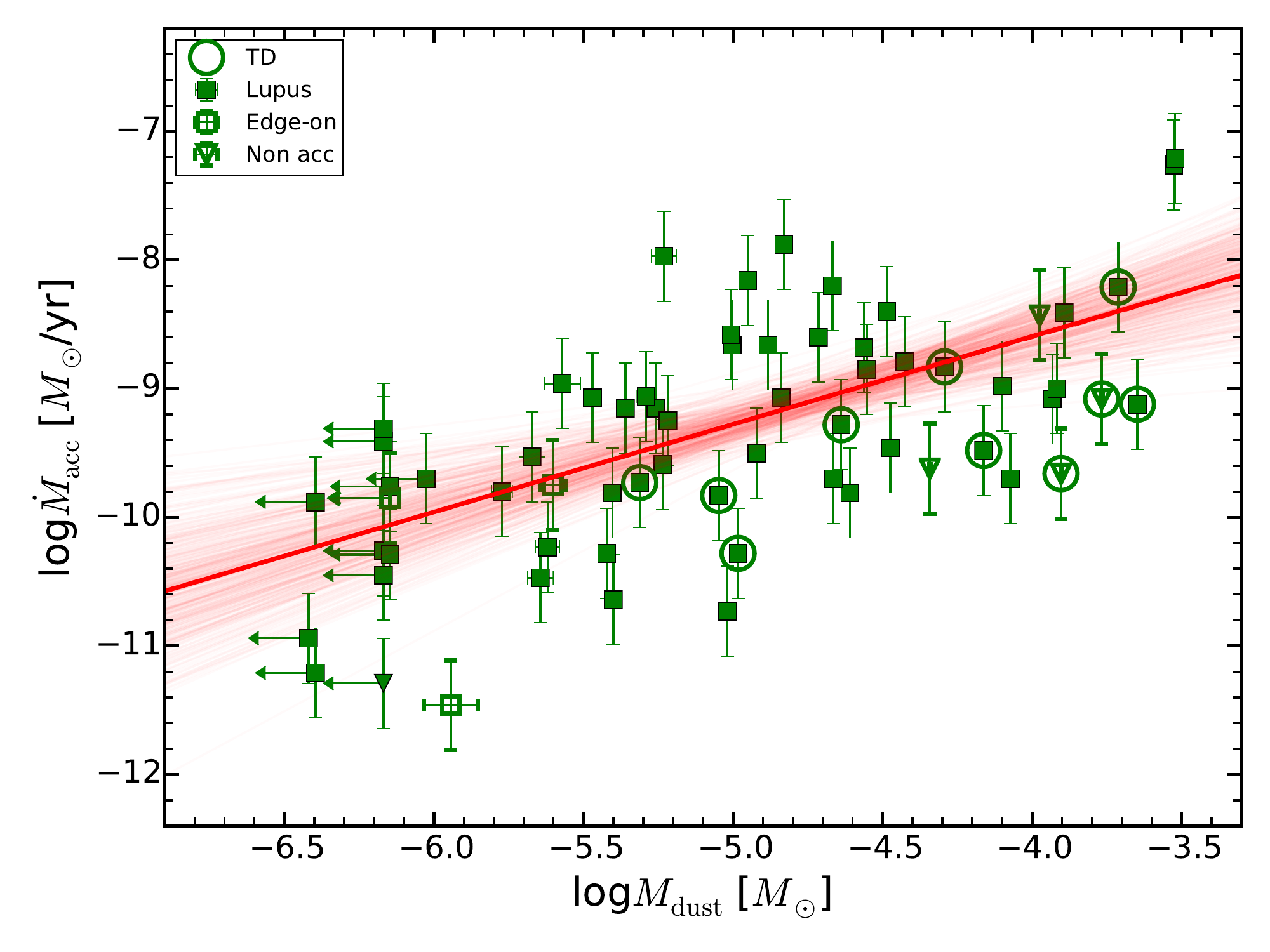}
\caption{Logarithm of \macc \ vs. logarithm of \mdust. Green filled squares are used for measured values, open squares for edge-on objects, and downward pointing open triangle for objects with accretion compatible with chromospheric noise.
Transition disks are indicated with a circle. 
We show fit results obtained using the Bayesian fitting procedure by \citet{Kelly07}, which considers errors on both axes and is only applied to detected targets. The assumed best fit is represented with a red solid line,   
while the light red lines are a subsample of the results of some chains. 
The best fitting with this procedure overlaps with the least-squares best fit. 
     \label{fig::macc_fmm}}
\end{figure}
%%%%%%%%%%%%%%%%%%%%%%%%%%%%%%%%%%%%%%%%%%%%%%%%%%%%%%%%%%%%%%%%%%%%%%%%%%%%

The values of \mdust \ derived by \citetalias{Ansdell16} (see Sect.~\ref{sect::sample}) are a measure of the bulk \textit{dust} mass in the disk. Figure~\ref{fig::macc_fmm} shows the values of \macc \ measured for our targets as a function of \mdust\footnote{Objects observed edge-on or with accretion compatible with chromospheric noise are shown in the plot, but they are not included in the analysis of correlations between \macc \ and \mdust.}. 

We first search for a correlation between the two quantities running a least-squares linear regression
on the targets with both \mdust \ and \macc \ measurements and find a moderate correlation with $r=0.53$ and a two-sided p-value of 1.5$\cdot 10^{-4}$ for the null hypothesis that the slope of this correlation is zero. The best fit obtained with this method has a slope of 0.7 and a standard deviation of the fit of 0.2. Then, we compute the linear regression coefficients using the fully Bayesian method by  \citet{Kelly07}\footnote{https://github.com/jmeyers314/linmix}, which allows us to include uncertainties on both axes in the fitting procedure. Uniform priors are used for the linear regression coefficients. 
We include some single chain results in Fig.~\ref{fig::macc_fmm}, as well as the best fit obtained with this method, which has the same slope and intercept of the least-squares fit relation. We adopt the median of the results of the chains as best fit values. We refer to Appendix~\ref{app::corner_plots} for the corner plots with the posterior analysis results. The best fit obtained with this method has a slope of $0.7 \pm 0.2$, 
a standard deviation of $0.4\pm0.1$, and a correlation coefficient of $0.56\pm0.12$. We also verified that the two quantities are still correlated when upper limits on \mdust \ are properly considered using the same tool. The correlation coefficient increases to $0.7\pm0.1$, while the slope is larger ($1.2\pm0.2$) but compatible with that obtained using detections only. The same slope is obtained including upper limits and using the emmethod and buckleyjames method in ASURV. However, the slope estimated when including upper limits is not well constrained \citep{Kelly07} and should be considered with caution. We then find a probability lower than $10^{-4}$ of no-correlation using the Cox hazard test for censored data in ASURV \citep{Lavalley92} including upper limits on \mdust. 

%%%%%%%%%%%%%%%%%%%%%%%%%%%%%%%%%%%%%
\begin{figure}[!t]
\centering
\includegraphics[width=0.5\textwidth]{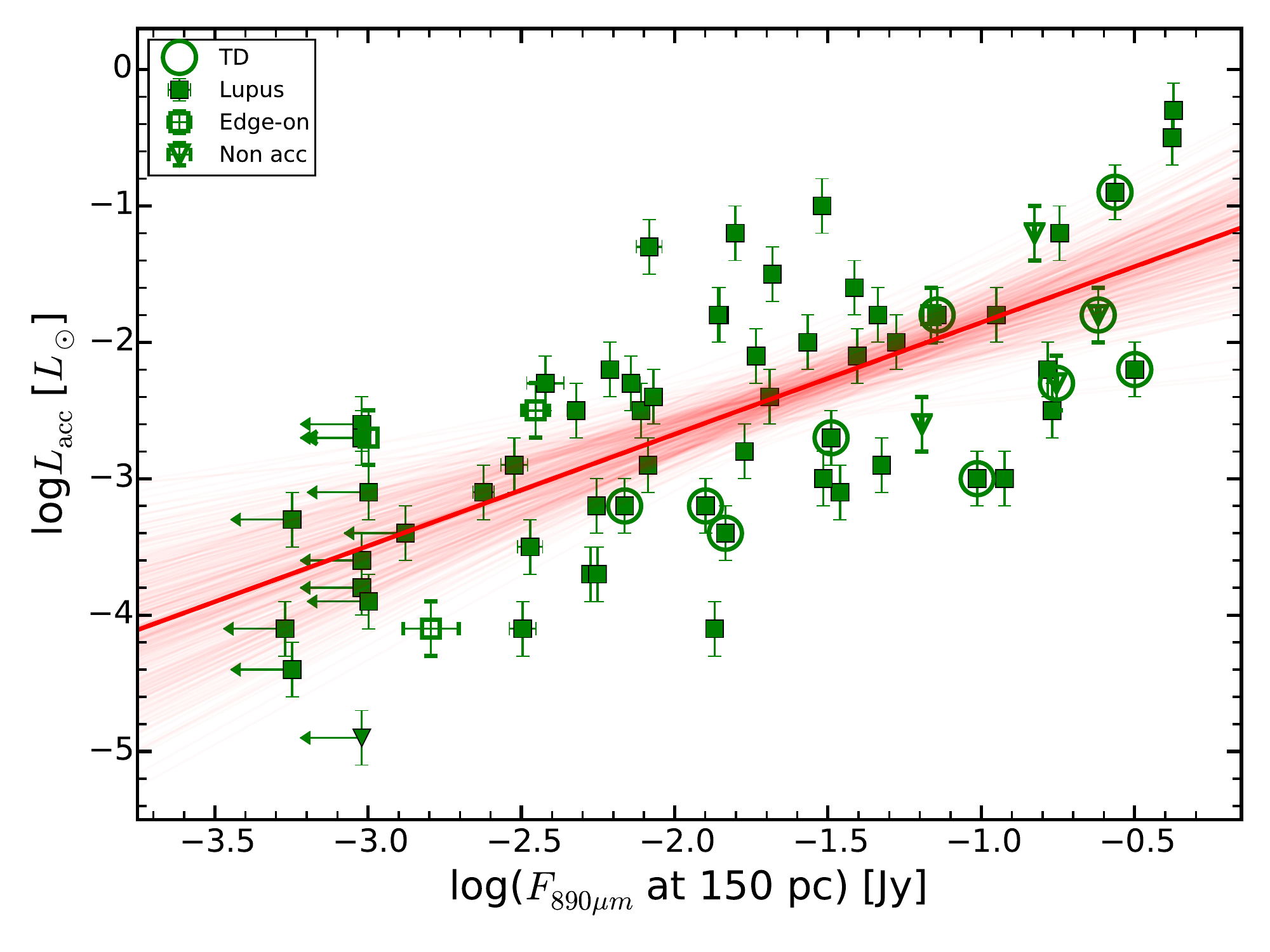}
\caption{Logarithm of \lacc \ vs. logarithm of continuum emission normalized to a distance of 150 pc. Symbols are as in Fig.~\ref{fig::macc_fmm}. 
     \label{fig::lacc_fmm}}
\end{figure}
%%%%%%%%%%%%%%%%%%%%%%%%%%%%%%%%%%%%%%%%%%%%%%%%%%%%%%%%%%%%%%%%%%%%%%%%%%%%

We show the dependence of the accretion luminosity (\lacc) on the sub-mm continuum flux normalized to a distance of 150 pc in Fig.~\ref{fig::lacc_fmm} to confirm that the correlation is not induced by the conversion from \lacc \ to \macc. Indeed, \lacc \ is directly measured from the spectra, while the conversion to \macc \ depends on \mstar, which is derived from evolutionary models. A correlation is found with r = 0.6, a slope of 0.8$\pm$0.2, and a standard deviation of 0.5$\pm$0.1.

We then test for the robustness of the correlation, given our assumptions to convert the continuum emission in \mdust. First, we assumed a single disk opacity and gas-to-dust ratio for all disks. To test whether a random variation of these parameters would affect our results, 
we perform the same statistical tests on the same targets after randomly displacing the values of \mdust \ within a uniform distribution with size $\pm$1 dex and centered on the measured value to mimic the uncertainties. We perform this test ten times, and the correlation is still present in nine out of ten realizations. 
Then, we test the effect of our assumption of a single $T_{\rm dust}$ by modifying our \mdust \ values assuming $T_{\rm dust}\propto L_\star^{0.25}$ \citep{Andrews10}. The correlation becomes less robust, but still significant ($r$=0.3, p-value=0.03).
We conclude that there is a statistically significant relation between the logarithm of \macc \ and the logarithm of \mdust. This relation has a slope slightly smaller than unity.

The location of TDs in Fig.~\ref{fig::macc_mdust} is also highlighted. All but one of the TDs are found to be below the best-fit relation in agreement with, for example, \citet{Najita15}. This suggests they have either lower \macc, or larger disk mass, or a different gas-to-dust ratio, than typical full disks.

\subsection{Disk gas mass}\label{sect::obs_macc_mgas}

%%%%%%%%%%%%%%%%%%%%%%%%%%%%%%%%%%%%%
\begin{figure}[!t]
\centering
\includegraphics[width=0.5\textwidth]{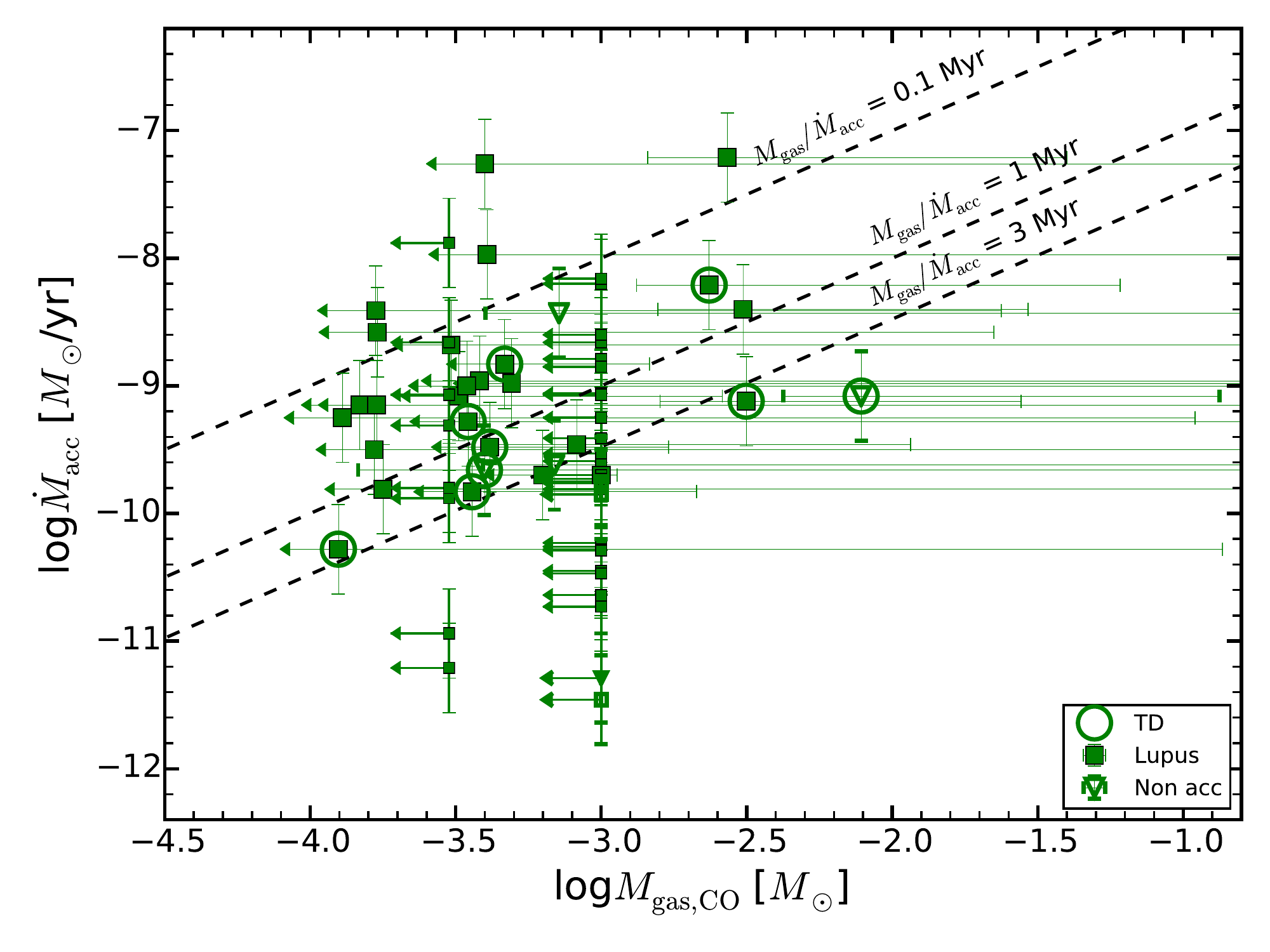}
\caption{Logarithm of \macc \ vs. logarithm of the disk mass derived from CO emission. Symbols are as in Fig.~\ref{fig::macc_fmm}. No correlation is found between these quantities.
     \label{fig::macc_mgas}}
\end{figure}
%%%%%%%%%%%%%%%%%%%%%%%%%%%%%%%%%%%%%%%%%%%%%%%%%%%%%%%%%%%%%%%%%%%%%%%%%%%%

The lower detection rates of CO lines than continuum emission \citepalias{Ansdell16} implies that we only measure \mgas \  for a few objects. Fig.~\ref{fig::macc_mgas} reports the \macc \ vs. \mgas \ plot. 

We perform the same statistical tests as for the \macc-\mdust \ relation. We find no correlation between the logarithm of \macc \ and the logarithm of \mgas \ using the least-squares linear regression on the targets with measured \mgas \ ($r=0.2$, p-value = 0.3). When considering uncertainties on the measurements we find a value for the correlation coefficient of 0.5$^{+0.4}_{-0.6}$, and thus we find no correlation. We obtain the same statistically insignificant value for the correlation coefficient, which points toward no correlation, even when we include the upper limits. Finally, the Cox hazard test for censored data gives a probability of 0.25 that the two quantities are not correlated. We then conclude that we do not detect any correlation between these two quantities. The large number of upper limits compared to detection is probably a limiting factor in studying this relation and the large error bars of the measurements are another limiting factor. Deeper ALMA surveys of CO emission in protoplanetary disks are needed to further study this relation.
In Fig.~\ref{fig::macc_mgas}, the TDs are mixed with full disks.

\section{Discussion} 

As mentioned in the Introduction, viscous evolution theory predicts that \macc$\propto$ $M_{\rm disk}/t_\nu(R_{\rm out})$. The evolution of the surface density of the disk ($\Sigma$) can be analytically described provided that the viscosity ($\nu$) is known \citep[e.g.,][]{Pringle81,Lodato08}.

As described by \citet{Jones12}, a similarity solution ($M_{\rm disk} \propto t^{-\sigma}$) is reached for times much larger than the viscous timescales under the simple assumptions $\nu \propto R^n$ or $\nu \propto \Sigma^m R^n$, where $R$ is the disk radius.
By differentiating this solution, one obtains $\dot{M}_{\rm acc} \propto \sigma t ^{-(1+\sigma)}$ and thus it is possible to define the ``viscous disk age'' as $t_{\rm disk}$ = \mdisk/\macc = $t/\sigma$. Measurements of the decline of \macc \ with time suggest that $\sigma\sim0.5$ \citep[e.g.,][]{Hartmann98,Sicilia-Aguilar10}. For viscously evolving disks, this implies that the age of the objects should be within a factor $\sim$2 of the ratio \mdisk/\macc. \citet{Jones12} have also shown that more complex assumptions on the disk viscosity, such as different values for the $\alpha$ viscosity description, lead to the same asymptotic behavior with \mdisk/\macc \ ratios usually larger than the age of the objects by a factor 2-3, but always less than 10. 

Other processes happening during disk evolution, such as a layered accretion, photoevaporation, and even planet formation, all lead to very similar values of \mdisk/\macc \ at late times, and these values are always higher than the age of the object \citep{Jones12}. Thus, a disk that evolved only from internal processes has a \mdisk/\macc \ ratio similar or larger than its age regardless of the assumption on the disk viscosity. The only means by which a disk might have an \mdisk/\macc \ ratio smaller than its age is if the disk is externally truncated (Rosotti et al., in prep.).

%%%%%%%%%%%%%%%%%%%%%%%%%%%%%%%%%%%%%
\begin{figure}[!t]
\centering
\includegraphics[width=0.5\textwidth]{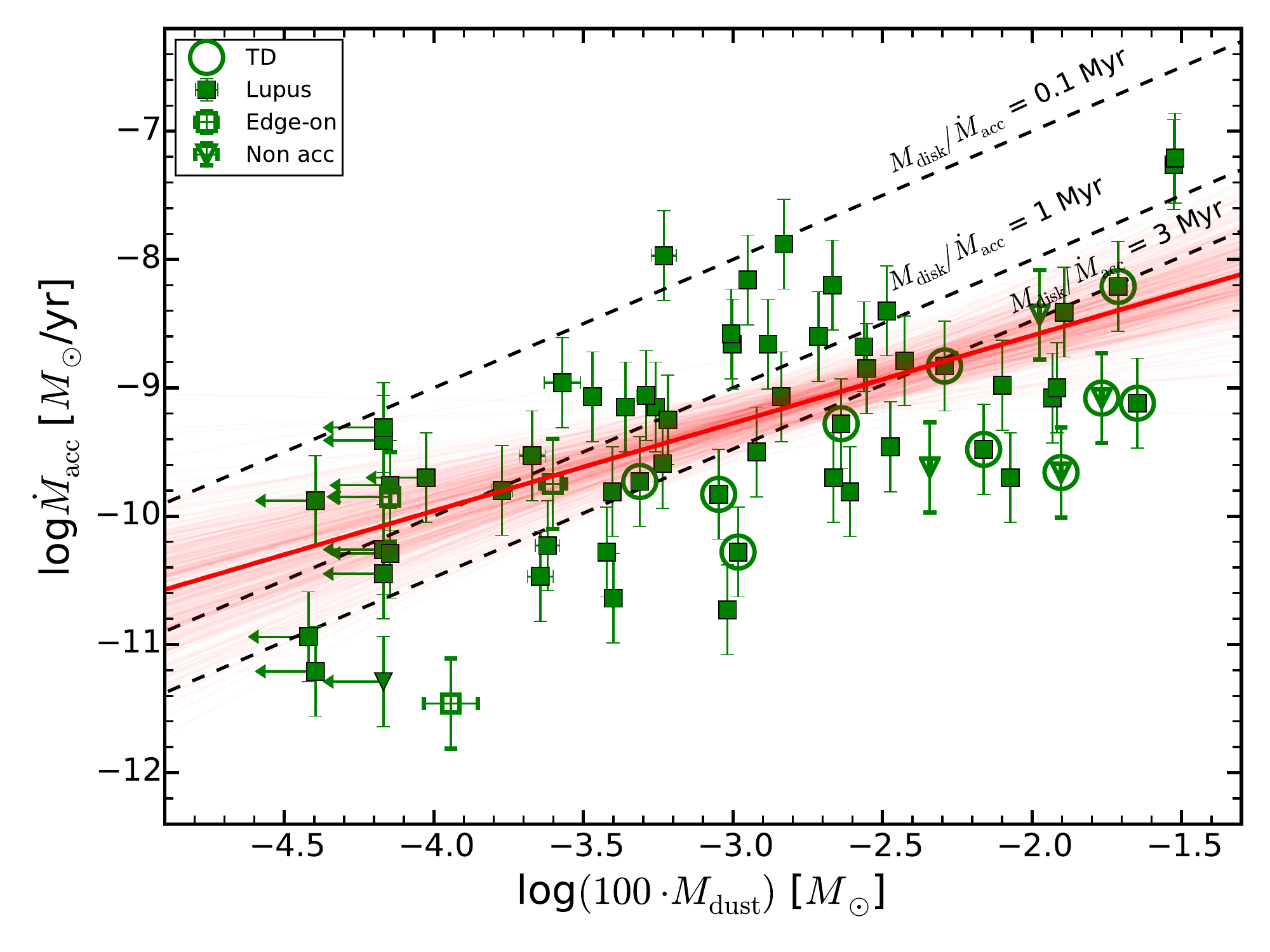}
\caption{Logarithm of \macc \ vs. logarithm of \mdisk = 100 $\cdot$ \mdust. 
Symbols are as in Fig.~\ref{fig::macc_fmm}. 
Also here, the best fit with the procedure by \citet{Kelly07} overlaps with the least-squares best fit. 
The dashed lines represent different ratios of \mdisk/\macc, as labeled.
     \label{fig::macc_mdust}}
\end{figure}
%%%%%%%%%%%%%%%%%%%%%%%%%%%%%%%%%%%%%%%%%%%%%%%%%%%%%%%%%%%%%%%%%%%%%%%%%%%%

We compare our results with these theoretical expectations by showing in Fig.~\ref{fig::macc_mgas} and~\ref{fig::macc_mdust} the \mdisk/\macc \ ratios for three different values of \mdisk/\macc = 0.1 Myr, 1 Myr, and 3 Myr. We assume that the \textit{total} disk mass (\mdisk) is \mdisk=\mgas~in Fig.~\ref{fig::macc_mgas}, while \mdisk=100$\cdot$\mdust \ in Fig.~\ref{fig::macc_mdust}. 
Indeed, to convert \mdust \ to \mdisk~ one needs to know the gas-to-dust ratio. We assume an interstellar medium (ISM) value of 100 for the gas-to-dust ratio, as is commonly done \citep[e.g.,][]{Andrews10,Ricci10}. If the gas-to-dust ratio has no dependence on \mdisk, this has no impact on the correlation between \macc \ and \mdisk \ but is instructive for the discussion. 
The typical age of Lupus targets is $\sim$1-3 Myr with a spread of 1-2 Myr \citep[e.g.,][]{Alcala14}. 

The location of the targets in Fig.~\ref{fig::macc_mdust} is in general agreement with the aforementioned theoretical expectations. Most of the targets (60\%) have positions between or compatible with the 1 and 3 Myr lines. 
However, several of the targets in Fig.~\ref{fig::macc_mgas} do not match the expectations from viscous evolution theory as they lie above the \mdisk/\macc = 1 Myr line.

The lack of correlation between \macc \ and \mgas \ is contrary to expectations from viscous evolution theory. When assuming \mdisk=100$\cdot$\mdust, however, we find  a correlation between \mdisk \ and \macc \ and also \mdisk/\macc \ ratios that are compatible with expectations from theory. Thus, we are inclined to conclude that the total disk mass \mdisk$\propto$\mdust, as with this assumption the correlation is present. This in turn suggests that \mgas \ measured from CO emission is possibly lower than the \textit{total} \mdisk, at least for the more massive disks. A possible explanation for this might be that carbon is processed in more complex molecules \citep[e.g.,][]{Bergin14,Kama16} or that more detailed modeling of CO lines is needed, but this discussion is out of the scope of this paper. 

The slope of the observed correlation between \macc \ and \mdisk, as measured from dust emission, is consistent with being linear, as expected if all disks evolve viscously, however,  we cannot exclude that it is actually shallower.
The exact slope can be derived with a better handle on the uncertainty in the \mdisk \ estimate, namely the gas-to-dust ratio, disk grain opacity, disk temperature, and their dependence on the stellar properties. More constraints on these values are awaited from future ALMA survey of disks with higher sensitivity, multiple band observations, and targeting several molecules in order to better determine the chemical properties of the disks.
The interest in further constraining this slope is related to the fact that this relation can tell us what evolutionary processes dominate at different stellar masses.

%__________________________________________________________________

\section{Conclusions}

In this Letter we compared the most complete and homogeneous datasets of properties of young stars and their disks to date. We used accretion rates onto the central star determined from UV excess with the VLT/X-Shooter spectrograph and disk masses from both sub-mm continuum and CO line emission measured by ALMA. 

We detected a statistically significant correlation between \macc \ and \mdisk \ with a slope that is slightly smaller than 1. This is found when assuming that the total disk mass is proportional to the disk dust mass, but not when using the disk gas mass. The latter result could be due to large uncertainties in \mgas \ estimate and low number statistics. For this reason, deeper surveys of gas emission in disks are needed. 
When measuring \mdisk \ from dust emission, transitional disks are found to have either a smaller \macc \ or a larger \mdisk \ than full disks.

We compared the observed \mdisk \ derived from dust emission and \macc \ with basic predictions from viscous evolution theory and we found a good agreement with the expected \mdisk/\macc \ ratios for our targets. 

Future studies should look for the \mdisk-\macc \ correlation for objects with different ages and in different environments.

\begin{acknowledgements}
      We thank Cathie Clarke and Phil Armitage for insightful discussions. We thank the anonymous referee for insightful comments that helped to improve the presentation of the results. CFM gratefully acknowledges an ESA Research Fellowship. GR is supported by the DISCSIM project, grant agreement 341137 funded by the European Research Council under ERC-2013-ADG. AN would like to acknowledge funding from Science Foundation Ireland (Grant 13/ERC/I2907). Leiden is supported by the European Union A-ERC grant 291141 CHEMPLAN, by the Netherlands Research School for  Astronomy (NOVA), and by grant 614.001.352 from the Netherlands Organization for Scientific Research (NWO). JPW and MA were supported by NSF and NASA grants AST-1208911 and NNX15AC92G.
\end{acknowledgements}


\appendix

\section{Sample}

\begin{table*}  
\begin{center} 
\footnotesize 
\caption{\label{tab::sample1} List of targets included in the analysis } 
\begin{tabular}{l| cc   } 
\hline \hline 
Object &  RA (J2000) & DEC (J2000)    \\  
% &   &   \\  
\hline 
1. 2MASSJ16100133-3906449 & 16:10:01.32 & -39:06:44.90 \\ 
2. SSTc2dJ154508.9-341734 & 15:45:08.88 & -34:17:33.70 \\ 
3. SSTc2dJ161243.8-381503 & 16:12:43.75 & -38:15:03.30 \\ 
4. MYLup$^{a}$ & 16:00:44.53 & -41:55:31.20 \\ 
5. SSTc2dJ160830.7-382827$^{a}$ & 16:08:30.70 & -38:28:26.80 \\ 
6. Sz68$^{a,c}$ & 15:45:12.87 & -34:17:30.80 \\ 
7. Sz133$^{b}$ & 16:03:29.41 & -41:40:02.70 \\ 
8. RYLup & 15:59:28.39 & -40:21:51.30  \\ 
9. Sz98 & 16:08:22.50 & -39:04:46.00 \\ 
10. SSTc2dJ160927.0-383628 & 16:09:26.98 & -38:36:27.60 \\ 
11. Sz118 & 16:09:48.64 & -39:11:16.90 \\ 
12. Sz73 & 15:47:56.94 & -35:14:34.80  \\ 
13. Sz83 & 15:56:42.31 & -37:49:15.50  \\ 
14. Sz90 & 16:07:10.08 & -39:11:03.50 \\ 
15. SSTc2dJ161344.1-373646$^{c}$ & 16:13:44.11 & -37:36:46.40 \\ 
16. Sz65$^{a,c}$ & 15:39:27.78 & -34:46:17.40 \\ 
17. Sz88A & 16:07:00.54 & -39:02:19.30  \\ 
18. Sz129 & 15:59:16.48 & -41:57:10.30 \\ 
19. Sz106$^{b}$ & 16:08:39.76 & -39:06:25.30 \\ 
20. 2MASSJ16085324-3914401 & 16:08:53.23 & -39:14:40.30 \\ 
21. Sz111 & 16:08:54.68 & -39:37:43.90 \\ 
22. SSTc2dJ161018.6-383613$^{c}$ & 16:10:18.56 & -38:36:13.00 \\ 
23. SSTc2dJ160026.1-415356$^{c}$ & 16:00:26.13 & -41:53:55.60 \\ 
24. Sz95$^{c}$ & 16:07:52.32 & -38:58:06.30 \\ 
25. Sz71 & 15:46:44.73 & -34:30:35.50  \\ 
26. Sz96 & 16:08:12.62 & -39:08:33.50 \\ 
27. Sz117 & 16:09:44.34 & -39:13:30.30 \\ 
28. SSTc2dJ160000.6-422158$^{c}$ & 16:00:00.62 & -42:21:57.50 \\ 
29. Sz131 & 16:00:49.42 & -41:30:04.10 \\ 
30. Sz72 & 15:47:50.63 & -35:28:35.40  \\ 
31. Sz123B$^{b}$ & 16:10:51.31 & -38:53:12.80  \\ 
32. Sz130$^{c}$ & 16:00:31.05 & -41:43:37.20 \\ 
33. Sz66$^{c}$ & 15:39:28.29 & -34:46:18.30 \\ 
34. 2MASSJ16085529-3848481$^{c}$ & 16:08:55.29 & -38:48:48.10 \\ 
35. Sz123A$^{c}$ & 16:10:51.60 & -38:53:14.00 \\ 
36. Sz74 & 15:48:05.23 & -35:15:52.80  \\ 
37. SSTc2dJ160002.4-422216 & 16:00:02.37 & -42:22:15.50 \\ 
38. Sz97 & 16:08:21.79 & -39:04:21.50 \\ 
39. Sz99 & 16:08:24.04 & -39:05:49.40 \\ 
40. Sz103 & 16:08:30.26 & -39:06:11.10 \\ 
41. Par-Lup3-3 & 16:08:49.40 & -39:05:39.30 \\ 
42. Sz110 & 16:08:51.57 & -39:03:17.70 \\ 
43. SSTc2d160901.4-392512 & 16:09:01.40 & -39:25:11.90 \\ 
44. Sz69$^{c}$ & 15:45:17.42 & -34:18:28.50 \\ 
45. Par-Lup3-4$^{b}$ & 16:08:51.43 & -39:05:30.40 \\ 
46. Sz113 & 16:08:57.80 & -39:02:22.70 \\ 
47. Sz115 & 16:09:06.21 & -39:08:51.80 \\ 
48. Sz88B & 16:07:00.62 & -39:02:18.10  \\ 
49. SSTc2dJ155925.2-423507 & 15:59:25.24 & -42:35:07.10 \\ 
50. 2MASSJ16081497-3857145 & 16:08:14.96 & -38:57:14.50 \\ 
51. Sz114 & 16:09:01.84 & -39:05:12.50 \\ 
52. AKC2006-19 & 15:44:57.90 & -34:23:39.50 \\ 
53. Sz104 & 16:08:30.81 & -39:05:48.80 \\ 
54. Sz112$^{c}$ & 16:08:55.52 & -39:02:33.90 \\ 
55. SST-Lup3-1 & 16:11:59.81 & -38:23:38.50 \\ 
56. Sz84 & 15:58:02.53 & -37:36:02.70  \\ 
57. Lup713$^{c}$ & 16:07:37.72 & -39:21:38.80 \\ 
58. Lup604s & 16:08:00.20 & -39:02:59.70 \\ 
59. Sz100$^{c}$ & 16:08:25.76 & -39:06:01.10 \\ 
60. Lup607$^{a,c}$ & 16:08:28.10 & -39:13:10.00  \\ 
61. Sz108B$^{c}$ & 16:08:42.87 & -39:06:14.70 \\ 
62. 2MASSJ16085373-3914367$^{c}$ & 16:08:53.73 & -39:14:36.70 \\ 
63. SSTc2dJ161029.6-392215 & 16:10:29.57 & -39:22:14.70 \\ 
64. Sz81A & 15:55:50.30 & -38:01:33.00  \\ 
65. SSTc2dJ161019.8-383607$^{c}$ & 16:10:19.84 & -38:36:06.80 \\ 
66. Lup818s$^{c}$ & 16:09:56.29 & -38:59:51.70 \\ 
\hline 
\end{tabular} 
\tablefoot{$^{a}$ Accretion compatible with chromospheric noise \citep{Alcala14,Alcala16}. $^{b}$ Edge-on targets \citep{Alcala14,Alcala16}. $^{c}$ Binary with separation less than 7$\arcsec$. } 
\end{center} 
\end{table*}  

\clearpage

\section{Corner plots of fit}\label{app::corner_plots}

\begin{figure*}[!t]
\centering
\includegraphics[width=0.7\textwidth]{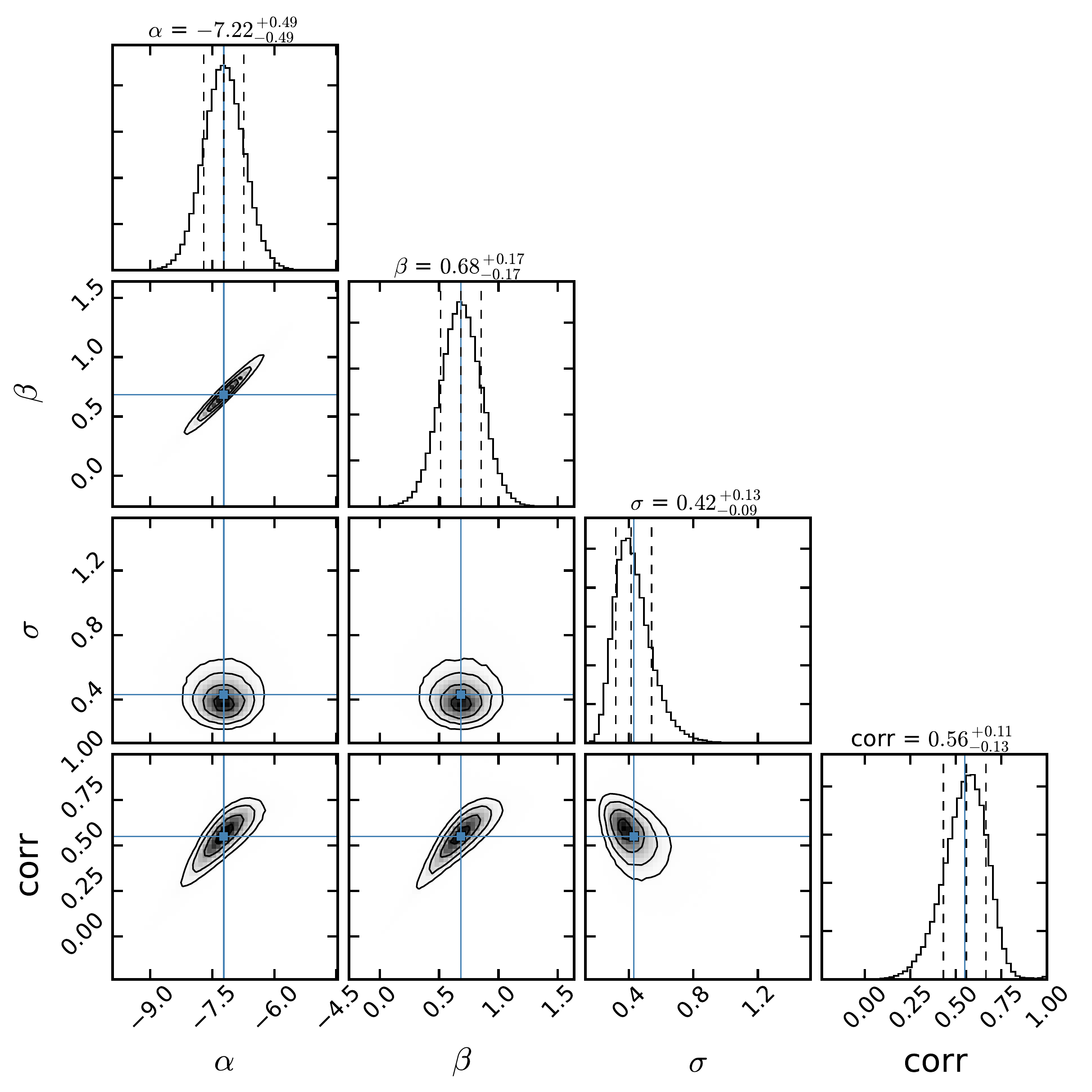}
\caption{Corner plot of the log\macc \ vs log(100$\cdot$\mdust) fit done using the linmix routine \citep{Kelly07}.
     \label{fig::corn_macc_mdust}}
\end{figure*}

We show in Fig.~\ref{fig::corn_macc_mdust} the corner plots showing the intercept, slope, standard deviation, and correlation coefficient for the fit of the log\macc \ vs log(100$\cdot$\mdust) relation (see Sect.~\ref{sect::obs_macc_mdust} and Fig.~\ref{fig::macc_mdust}). This is done using linmix  \citep{Kelly07}, only data with detection on both axes, and including the errors on both axes. 

%%%%%%%%%%%%%%%%%%%%%%%%%%%%%%%%%%%%%
%%%%%%%%%%%%%%%%%%%%%%%%%%%%%%%%%%%%%%%%%%%%%%%%%%%%%%%%%%%%%%%%%%%%%%%%%%%%

\end{document}